\documentclass[runningheads]{llncs}
\usepackage[T1]{fontenc}
\usepackage{url}
\usepackage{graphicx}
\begin{document}
\title{Perceived Femininity in Singing Voice: Analysis and Prediction}
\titlerunning{Perceived Femininity in Singing Voice: Analysis and Prediction}
%
\author{Yuexuan Kong \and
Viet-Anh Tran \and
Romain Hennequin}
\authorrunning{Y. Kong et al.}
%
\institute{Deezer Research, Paris, France\\
\email{ykong@deezer.com}}
\maketitle              


\begin{abstract}
This paper focuses on the often-overlooked aspect of perceived voice femininity in singing voices. While existing research has examined perceived voice femininity in speech, the same concept has not yet been studied in singing voice. The analysis of gender bias in music content could benefit from such study. To address this gap, we design a stimuli-based survey to measure perceived singing voice femininity (PSVF), and collect responses from 128 participants. Our analysis reveals intriguing insights into how PSVF varies across different demographic groups. Furthermore, we propose an automatic PSVF prediction model by fine-tuning an x-vector model, offering a novel tool for exploring gender stereotypes related to voices in music content analysis beyond binary sex classification. This study contributes to a deeper understanding of the complexities surrounding perceived femininity in singing voices by analyzing survey and proposes an automatic tool for future research.
\end{abstract}

\section{Introduction}


Recently, in the field of music content analysis, numerous studies have delved into social stereotypes and biases surrounding gender and sex, notably regarding lyrics \cite{betti2023large,shakespeare2020exploring,Smiler2017FromW}. From 1960 to 2010, popular song lyrics of male solo artists became more sexist over the years, while this behavior is less noticeable for the other categories of artists\cite{betti2023large}. Sexual behaviors and objectification in lyrics of both female and male are more frequently linked to male artists, and lyrics of male artists contain more gender bias than those of female artists\cite{betti2023large,Smiler2017FromW}. 

While these studies employ the concept of gender, they primarily frame the problem by associating content with the singer's biological sex. Yet, gender and sex are fundamentally different; gender being a socially constructed notion that may or may not align with one's biological sex.  More importantly, recent scientific consensus defines the notion of gender more as a continuous variable than a discrete one\cite{reilly2019gender}. Furthermore, in social psychology, on top of gender, perceived masculinity and femininity by humans have often been considered an important component while investigating the influence on social behaviors \cite{chen2023acoustic}. 

In speech, there has been a surge in research exploring the relationship between perceived voice femininity and social impressions. A previous research points out how perceived gender in voices is a socially and culturally constructed concept\cite{eidsheim2019race}. Some studies focus on social stereotypes of voices of different femininity \cite{ko2009stereotyping,arnocky2018men}; while others develop tools to predict perceived voice femininity in speech, aiding transgender individuals in their gender affirming surgeries \cite{doukhan2023voice,chen2023acoustic}.

Despite advancements in perceived voice femininity in speech, research on perceived femininity in singing voice is lacking. There are various research on gender estimation from singing voice \cite{hamasaki2014songrium,weninger2011combining,weninger2011automatic}. However currently, there is no data analysis of perceived singing voice femininity or model capable of predicting the average perceived singing voice femininity (PSVF), despite its potential usefulness. In speech, research shows that the perception of leadership capacity is influenced by pitch in voices\cite{klofstad2012sounds}, which is often related to perceived femininity\cite{munson2007acoustic}. In singing voice, similarly to speech, the message carried by the lyrics can be biased by the perceived femininity of the singing voice, rather than the biological sex of the singer when it is unknown to listeners. Thus having access to the information of PSVF can be of great help for sociological studies about gender representation in music and in media. This approach offers a broader perspective on voice characteristics compared to binary classification of sex often used in music content analysis. Moreover, introducing an automatic model for PSVF prediction could facilitate the analysis of bias, since collecting PSVF of songs for a large catalog is time consuming.

In this work, as a first step towards PSVF analysis, we have three main contributions:
\begin{itemize}

\item We design a stimuli-based PSVF survey, 
consisting of 1200 audio segments lasting 3 seconds each, balanced across five languages, four age groups, and two sexes, with 7258 valid responses gathered from 128 participants. The dataset is publicly available at \url{https://github.com/deezer/perceived_singing_voice_femininity}.

\item We use the survey responses to compare and analyze results among different groups of singers and participants, to test certain hypothesis on PSVF and gain a better understanding.

\item Finally, we use the dataset to fine-tune a modified x-vector model to perform regression instead of binary classification of PSVF, paving the way for large-scale analysis of biases in music corpora coupled to PSVF.

\end{itemize}

\section{Related Work}

The first research on automatic perceived voice femininity prediction in speech is conducted by using classification methods\cite{williamsapplication}. They use a Multi-Layer Perceptron (MLP) for transgender self-assessment of voice femininity. This model is trained using binary gender data and flexible thresholds to categorize voices into masculine, feminine, and androgynous classes. Its accuracy, measured against speaker self-assessments, stands at 88\%. However, this system cannot describe perceived femininity on a continuous scale. Chen et al. \cite{chen2020objective,chen2023acoustic} used Linear Discriminant Analysis(LDA) to align the results of their model with human perception of voice on a continuous scale and conducted an analysis of acoustic factors that impact human perception. To assist transgender individuals in voice training following gender-affirming treatments, Doukhan et al.~\cite{doukhan2023voice} conducted a survey of human perception of transgender voices and trained a binary classification system based on x-vector\cite{snyder2018x}. They then used isotonic regression calibration procedure to transform the results into a continuous scale to predict the perceived femininity of transgender individuals\cite{chakravarti1989isotonic}. To predict the voice femininity percentage, they deployed an x-vector architecture which is a time-delayed neural network that embeds voice characteristics. 

However, to the best of our knowledge, due to the lack of data, singing voice has not been studied in the context of perceived femininity analysis prediction.

\section{Human Perception of Singing Voice Femininity}
\label{human performance}
In this section, we conduct a detailed evaluation of human perception of singing voice's femininity, with the aim of enhancing our comprehension of perceived singing voice femininity (PSVF) and differences in perceptions towards various groups of singers (language, sex, age). We use the term voice femininity as previous research on voice femininity in speech\cite{doukhan2023voice,doukhan2018describing}.

\subsection{Survey Design and Data Collection}
In our study, we use the test set of STraDa\cite{strada}. STraDa contains 200 songs that are distributed equally across two sexes of the lead singer, 4 age groups of the lead singer (20-34, 35-49, 50-64, 65+) and 5 languages (Mandarin, English, French, Spanish, German), whose collection process is detailed in STraDa \cite{strada}. We refer these groups as subgroups of singers. All songs are annotated by the authors manually, and only cis-gender singers are used, which means their gender corresponds to their biological sex. To increase the number of segments of survey, we choose 6 segments of 3 seconds from each song that contain vocals.

The 1200 segments are randomly divided into 20 surveys, each containing 60 segments. The surveys are administered through the online platform \textit{JotForm}\footnote{https://www.jotform.com}. We ask each participant to rate the perceived femininity of the corresponding singing voice in the segment. Each segment is rated using a Likert scale, with the following values corresponding from 2 to -2: "definitely feminine", "rather feminine", "I don't know", "rather masculine" and "definitely masculine". Positive values correspond to feminine and negative values correspond to masculine. We collect participants' metadata (gender, language, and age group) at the beginning of the survey. We refer these groups as subgroups of participants. Additionally, after each question, participants are asked if they recognize the singer in the corresponding segment, to eliminate cases where participants' perceptions are biased by prior knowledge. At the end of the survey, we ask participants whether they experienced any difficulties during the listening task. We share this survey through community mailing lists and in several universities in France and China, and obtain 128 participants. We remove responses indicating prior knowledge of the singer, two participants who have reported difficulties during the task were excluded, resulting in 126 participants and 7258 valid responses for 1200 segments. We release these data to facilitate future research on PSVF\footnote{\url{https://github.com/deezer/perceived_singing_voice_femininity}}.

To summarize the survey results and analyze the PSVF, as well as to compare perceptions among different subgroups of segments and participants, we introduce the term \textit{average correspondence (AC)}. This is calculated as the percentage of instances within each subgroup where the averaged PSVF across all participants aligns with the singer's biological sex. For example, within the subgroup of female singers' segments, the average PSVF was positive in 96.7\% of all segments, indicating a perception of femininity that aligns with the singers' sex. It serves as a metric to indicate whether the singing voices within one subgroup differ more from the stereotypical perception of voices associated with sex compared to another subgroup.

\subsection{Comparison of Different Subgroups' PSVF}
\label{human biases}

Our hypothesis is that subgroups might exhibit higher AC for singing voices they are more accustomed to or that are physiologically similar to their own (e.g., french participants to french songs). However, Table \ref{AC} reveals no significant differences in AC among the various participant subgroups. This suggests that factors such as gender, language, and age do not significantly influence the AC of participants for different subgroups of singers in our study. All subgroups with different gender/age/language exhibit higher AC for male singers, singers of age group 35-49 and mandarin tracks. These findings could be helpful in future research, as they suggest that participants' PSVF may not be strongly influenced by demographic factors, and may be more universally applicable.

\begin{table}[h]
\centering

\begin{minipage}[t]{0.48\linewidth}
\centering
\begin{tabular}{ccc}
\hline
 & \textbf{female} & \textbf{male} \\\hline 
 \textbf{nb. participants} & 53 & 73 \\ \hline
 \textbf{AC female singers} & 82.9 & 87.9 \\
 \textbf{AC male singers} & \textbf{96.7} & \textbf{96.4} \\\hline
\end{tabular}
\end{minipage}
\hfill
\begin{minipage}[t]{0.48\linewidth}
\centering
\begin{tabular}{cccc}
\hline
 & \textbf{20-34} & \textbf{35-49} & \textbf{50-65} \\\hline 
 \textbf{nb. participants} & 96 & 24 & 4 \\ \hline
 \textbf{AC 20-34 singers} & 96.3 & 93.3 & 87.2 \\
 \textbf{AC 35-49 singers} & \textbf{97.0} & \textbf{97.7} & \textbf{94.9} \\
 \textbf{AC 50-64 singers} & 87.6 & 82.2 & 81.0 \\
 \textbf{AC 65+ singers} & 92.5 & 87.4 & 81.8 \\\hline
\end{tabular}
\end{minipage}

\vspace{1em}

\begin{minipage}[h!]{\linewidth}
\centering
\begin{tabular}{cccccc}
\hline 
 & \textbf{fr} & \textbf{en} & \textbf{sp} & \textbf{man} & \textbf{ge} \\\hline 
 \textbf{nb. participants} & 67 & 108 & 15 & 24 & 8 \\ \hline
 \textbf{AC fr tracks} & 85.8 & 87.9 & 89.6 & 85.1 & 85.9 \\
 \textbf{AC en tracks} & 92.0 & 93.3 & 93.0 & 91.4 & 93.2 \\
 \textbf{AC sp tracks} & 90.0 & 89.6 & 80.9 & 87.4 & 85.9 \\
 \textbf{AC man tracks} & \textbf{97.7} & \textbf{97.9} & \textbf{95.9} & \textbf{92.0} & 94.3 \\
 \textbf{AC ge tracks} & 92.7 & 94.6 & 92.1 & 89.8 & \textbf{94.7} \\\hline
\end{tabular}
\end{minipage}


\caption{AC (\%) of different subgroups. Answers of participants older than 65 years old do not cover all age groups, therefore we do not report the ACs.}
\label{AC}
\end{table}

Furthermore, to compare uncertainty across different subgroups, we calculate the proportion of segments where the averaged answer falls between -0.5 and +0.5 for each subgroup of singers (denoted as \textit{Unsure}). The results are shown in Table \ref{table:unsure}.

\begin{table}[h!]
\centering
\begin{tabular}{c|cc|cccc|ccccc}
\hline
 & \multicolumn{2}{c|}{\textbf{Gender}} & \multicolumn{4}{c|}{\textbf{Age}} & \multicolumn{5}{c}{\textbf{Language}} \\
 & female & male & 20--34 & 35--49 & 50--64 & 65+ & fr & en & sp & man & ge \\
\hline
\textbf{Unsure (\%)} & \textbf{6.8} & 2.3 & 2.3 & 0.3 & \textbf{9.1} & 4.1 & \textbf{7.1} & 2.1 & 5.4 & 0.8 & 4.5 \\
\hline
\end{tabular}

\caption{\textit{Unsure} (\%) of different subgroups of singers.}
\label{table:unsure}
\end{table}

Participants perceive female voices, voices of singers aged 50-64, and French singing voices as more neutral, represented by higher \textit{Unsure} in Table \ref{table:unsure}. This suggests that on average, participants find these voices less stereotypical compared to voices typically associated with both sexes. While we acknowledge that observations of \textit{Unsure} may partly result from sampling bias in STraDa, we would like to shed more lights in potential reasons for these similarities beyond sampling bias. For singer's sex, female voices might have become less stereotypical over the time, with the increase of representation of female singers in genres such as blues and gospel singing, where female singers tend to have deeper and more powerful voices that were more typically associated with male voices\cite{griffin2004malindy}. For language, music in Mandarin is known to differ from Western music in tonality, rhythm, and melody~\cite{rahn1999chinese}, which can impact how individuals perceive femininity in singing voices. Additionally, traditional Chinese folk songs typically have higher-pitched female vocals, which have characteristics that resemble more to stereotypical voices that people perceive as feminine voice. For age groups, it may be due to changes in the human voice that occur with age. As individuals age, respiratory changes result in less efficient air movement, and larynx changes reduce vocal fold adjustments during voice production~\cite{hirano1989ageing,tarafder2012aging}. This could explain lower AC for older age groups where the fundamental frequency of female voices become closer to the fundamental frequency of male voices\cite{stathopoulos2011changes} and that people perceive these voices as more neutral. 

\subsection{Comparison of Human Perception and Machine's Singer Sex Classification Results}
By using the training set of STraDa\cite{strada}, the authors trained a singer sex classifier and evaluated it on the testing set of STraDa, the same dataset used for the survey. We compare our AC results with the findings from \cite{strada} to justify our decision to fine-tune the automatic model from \cite{strada} for large-scale catalog labeling. 

The subgroups that exhibit higher AC for PSVF and where the automatic evaluation system renders higher \textit{recall} in \cite{strada} are respectively the same. While this could be explained by sampling bias in the testing set of STraDa, we find that only 41 out of 114 cases are the same where PSVF and automatic sex classification system do not match with the singer's sex. This suggests that sampling bias alone could not explain the reasons behind this similarity.

This similarity also shows that potentially by fine-tuning this x-vector model, we could obtain a model that align its automatic prediction with PSVF.

\section{Automatic Perceived Singing Voice Femininity Prediction System}
Within the context of sociological analysis, where PSVF serves as a factor that broadens the perspective for analyzing biases present in music content such as lyrics, we require an automatic prediction model to label a large catalog.

In this section, we fine-tune on the x-vector model used in \cite{strada} to align it with PSVF. Moreover, in speech, Doukhan et al. also deployed a system based on x-vector to predict the perceived femininity which showed such system's potential in predicting perceived femininity\cite{doukhan2023voice}.

\subsection{Training Details}
We employ the identical set of 1200 segments with human annotations we obtain from the survey. It is worth noting that a binary coding is insufficient to fully capture PSVF, which is more of a continuous variable than a binary one\cite{doukhan2023voice}. We calculate the average of all responses for each one of the 1200 segments. Since the scale used in the survey is from -2 to 2, we rescale each answer and obtain a value that ranges from 0 to 1, as the reference for the automatic PSVF prediction system. We use a 5-fold cross-validation technique to assess the performance of the automatic PSVF prediction system, resulting in 860 segments for training and 240 segments for testing for each fold.

The architecture used in \cite{strada} is a x-vector embedding model. It contains 5 time-delayed neural network (TDNN) blocks and outputs an embedding of 64 dimensions, then a linear layer that projects embeddings into two classes. It outputs two neurons for sex classification, thus we replace the last layer by one neuron that outputs a value from 0 to 1 for a continuous prediction. We freeze the first two blocks of the TDNN and retrain other layers by using our training data. We use L1 loss to optimize the model.

We adapt all pre-processing steps used in \cite{strada}. We use mel-spectrograms using 24 Mel filters, as input for the TDNN model to extract x-vector embeddings. To improve the quality and quantity of our training data, we integrate source separation using Spleeter\cite{spleeter2020} for half of training samples and speed change\cite{yang2022torchaudio} as a method of data augmentation. 

\subsection{Results}

We use Mean Average Error (MAE) to measure the performance of our model, which is a direct indicator of how far the prediction falls from the survey result. Our model shows an average MAE of 0.10 with a standard deviation of 0.01 across five folds, indicating that, on average, our model's predictions deviate by 0.1 from the results obtained from the survey on a scale of 0 to 1. Considering the relatively limited size of our training data, this performance stands as a first step towards an automatic PSVF prediction model. We firmly believe that employing this model has the potential to offer a multifaceted lens through which biases in music content can be examined. It diverges from the conventional approach of studying biases related to sex, allowing for a broader exploration of perspectives within the realm of music content analysis.

\section{Conclusion}
In this paper, we adapt a concept that is recently introduced into speaking voice into the field of singing voice: perceived voice femininity. It shifts away from binary classification based on biological sex, and moves towards a richer understanding of how singing voices are gradually perceived in terms of femininity on a continuous scale. In doing so, our goal is to enable large-scale analysis of music corpora in terms of gender stereotypes, for instance by linking lyrics to perceived singing voice femininity (PSVF) of lead singers.

To achieve such goal, we create a stimuli-based survey to investigate PSVF of humans and will publish all related data for future use. We then analyze PSVF data to show how PSVF of different singers vary across different subgroups of participants. Trained on this data, we provide the community an automatic prediction tool that could be used on a large music catalog to conduct sociologist research related to gender stereotypes.

This study fills in the gap in research on analysis of PSVF, yet there is room for improvement. Firstly, future research could enhance the survey by adding singers of more diverse gender identities and gathering data from a more diverse range of individuals, thus aiming for a more representative estimation of PSVF. Additionally, influences of hyperparameter choices, model choices and segment lengths for automatic model could be investigated further. Lastly, investigating the underlying reasons why certain subgroups' singing voices diverge more from stereotypical voices could bring a better understanding of human perception of singing voice femininity.

\bibliographystyle{splncs04}
\bibliography{CMMR2025_LaTeX2e_Proceedings_Templates/cmmr2025_template_samplepaper}

\end{document}